\documentclass{llncs} %
\usepackage{float}
\usepackage{graphicx}
\usepackage{wrapfig}

\usepackage[labelfont=bf]{caption}
\usepackage{color}
\usepackage{amsmath} % Mathe
\usepackage{mathtools} % Mathe
\usepackage{amsfonts} % Mathesymbole
\usepackage{amssymb}
\usepackage[normalem]{ulem}

\usepackage{array}
\usepackage{booktabs}
\usepackage{rotating}
\usepackage{multirow}
\usepackage{multicol}
\usepackage{lscape}
\usepackage{subfig}
\usepackage{comment}
\usepackage[export]{adjustbox}

\usepackage{algorithm,algorithmicx,algpseudocode}

\usepackage{xcolor,soul}
\definecolor{pred}{HTML}{ffe699}
\definecolor{gt}{HTML}{d5c3d2}

\usepackage[colorlinks=true,linkcolor=blue,citecolor=blue]{hyperref}

\usepackage{tikz,xcolor,hyperref}

\definecolor{lime}{HTML}{A6CE39}
\DeclareRobustCommand{\orcidicon}{
	\begin{tikzpicture}
	\draw[lime, fill=lime] (0,0) 
	circle [radius=0.16] 
	node[white] {{\fontfamily{qag}\selectfont \tiny ID}};
	\draw[white, fill=white] (-0.0625,0.095) 
	circle [radius=0.007];
	\end{tikzpicture}
	\hspace{-2mm}
}

\foreach \x in {A, ..., Z}{\expandafter\xdef\csname orcid\x\endcsname{\noexpand\href{https://orcid.org/\csname orcidauthor\x\endcsname}
			{\noexpand\orcidicon}}
}
			
\definecolor{darkgreen}{rgb}{0.53, 0.66, 0.42}

\begin{document}

\title{Deep EvoGraphNet Architecture For Time-Dependent Brain Graph Data Synthesis From a Single Timepoint}

\titlerunning{Short Title}  % abbreviated title (for running head)

\author{Ahmed Nebli$^{\dagger}$ \orcidB{}\inst{1,2} \and U\u{g}ur Ali Kaplan$^{\dagger}$\inst{1} \and Islem Rekik\orcidA{} \index{Rekik, Islem}\inst{1}\thanks{ {corresponding author: irekik@itu.edu.tr, \url{http://basira-lab.com}; $^{\dagger}$: co-first authors.} This work is accepted for publication in the PRedictive Intelligence in MEdicine (PRIME) workshop Springer proceedings in conjunction with MICCAI 2020. }}

\institute{$^{1}$ BASIRA Lab, Faculty of Computer and Informatics, Istanbul Technical University, Istanbul, Turkey \\ $^{2}$ National School for Computer Science (ENSI), Mannouba, Tunisia }

\authorrunning{A. Nebli et al.}

\maketitle              % typeset the title of the contribution

\begin{abstract}

Learning how to predict the brain connectome (i.e. graph) development and aging is of paramount importance for charting the future of within-disorder and cross-disorder landscape of brain dysconnectivity evolution. Indeed, predicting the \emph{longitudinal} (i.e., \emph{time-dependent}) brain dysconnectivity as it emerges and evolves over time from a single timepoint can help design personalized treatments for disordered patients in a very early stage. Despite its significance, evolution models of the brain \emph{graph} are largely overlooked in the literature. Here, we propose EvoGraphNet, the first \emph{end-to-end} geometric deep learning-powered \emph{graph}-generative adversarial network (gGAN) for predicting time-dependent brain graph evolution from a single timepoint. Our EvoGraphNet architecture cascades a set of time-dependent gGANs, where each gGAN communicates its predicted brain graphs at a particular timepoint to train the next gGAN in the cascade at follow-up timepoint. Therefore, we obtain each next predicted timepoint by setting the output of each generator as the input of its successor which enables us to predict a given number of timepoints using only one single timepoint in an end-to-end fashion. At each timepoint, to better align the distribution of the predicted brain graphs with that of the ground-truth graphs, we further integrate an auxiliary  Kullback-Leibler divergence loss function. To capture time-dependency between two consecutive observations, we impose an $l_1$ loss to minimize the sparse distance between two serialized brain graphs. A series of benchmarks against variants and ablated versions of our EvoGraphNet showed that we can achieve the lowest brain graph evolution prediction error using a single baseline timepoint. Our EvoGraphNet code is available at \url{http://github.com/basiralab/EvoGraphNet}.


\end{abstract}

%% ***************************************************************************** %%
\section{Introduction}
%% ***************************************************************************** %%

Recent findings in neuroscience have suggested that providing personalized treatments for brain diseases can significantly increase the chance of a patient's recovery \cite{mukherjee2020}. Thereby, it is vital to undertake an early diagnosis of brain diseases \cite{lohmeyer2020}, especially for neurodegenerative diseases such as dementia which was found to be irreversible if discovered at a late stage \cite{Stoessl2012}. In this context, recent landmark studies \cite{Rekik:2017c,Gafuroglu:2018} have suggested using the robust predictive abilities of machine learning to predict the time-dependent (i.e., longitudinal) evolution of both the healthy and the disordered brain. However, such works only focus on predicting the brain when modeled as an image or a surface, thereby overlooking a wide spectrum of brain dysconnectivity disorders that can be pinned down by modeling the brain as a graph (also called connectome) \cite{van2019cross}, where the connectivity weight between pairs of anatomical regions of interest (ROIs) becomes the feature of interest.

In other works, \cite{li2019,liu2020} have proposed to use deep learning frameworks aiming to utilize hippocampal magnetic resonance imaging (MRI) to predict the onset of Alzheimer's disease (AD). However, these studies have only focused on a single brain region overlooking other brain regions' engagement in predicting AD. To overcome this limitation, \cite{zhang2015,islam2017} proposed to use brain MRI to predict AD across all brain regions, yet such experiments focused on MRI samples collected at the late stages of the illness which are inadequate for administering patient's personalized treatments at \emph{baseline observation}. Here, we set out to  solve a more difficult problem which is forecasting the evolution of the brain graph over time from an initial timepoint. Despite the lack of studies in this direction, \cite{bahapaper} has attempted to solve this challenge by suggesting the Learning-guided Infinite Network Atlas selection (LINAs) framework, a first-ever study that developed a learning-based sample similarity learning framework to forecast the progression of brain disease development over time from a single observation only. This study reflects a major advancement in brain disease evolution trajectory prediction since it considered using the brain as a set of interconnected ROIs, instead of performing image-based prediction which is agnostic to the nature of the complex wiring of the brain as a graph.

Regardless of the fact that the aforementioned techniques were aiming to address the issue of early diagnosis of brain diseases, all of these approaches share the same \emph{dichotomized} aspect of the engineered learning framework, each composed of independent blocks that cannot co-learn together to better solve the target prediction problem.  For instance, \cite{bahapaper} first learns the data manifold, then learns how to select the best samples before performing the prediction step independently. These sub-components of the framework do not interact with each other and thus there is no feedback passed on to earlier learning blocks.

To address these limitations and while drawing inspiration from the compelling generative adversarial network (GAN) architecture introduced in \cite{goodfellow2014}, we propose EvoGraphNet, a framework that first generalizes the GAN architecture, originally operating on Euclidean data such as images, to non-Euclidean graphs by designing a \emph{graph-based} GAN (gGAN) architecture. Second and more importantly, our  EvoGraphNet chains a set of gGANs, each specialized in learning how to generate a brain graph at follow-up timepoint from the predicted brain graph (output) of the previous gGAN in the \emph{time-dependent} cascade.  We formalize our brain graph prediction task from baseline using a single loss function to optimize with a backpropagation process throughout our EvoGraphNet architecture, trained in an end-to-end fashion. Our framework is inspired by the works of \cite{pix2pix2016,welander2018} where we aim to perform an \emph{assumption free} mapping from an initial brain graph to its consecutive time-dependent representations using a stack of $m$ paired  generators and discriminators at $m$ follow-up timepoints. Each generator inputs the output of its predecessor generator making the framework work in an end-to-end fashion.  We further propose to enhance the quality of the evolved brain graphs from baseline by maximizing the alignment between the predicted and ground-truth brain graphs at each prediction timepoint by integrating a Kullback-Leibler (KL) divergence.  To capture time-dependency between two consecutive observations, we impose an $l_1$ loss to minimize the sparse distance between two serialized brain graphs. We also explore the effect of adding a graph-topology loss where we enforce the preservation of the topological strength of each ROI (graph node) strength over time.

Below, we articulate the main contributions of our work: 

\begin{enumerate}

\item \emph{On a conceptual level.} EvoGraphNet is the first geometric deep learning framework that predicts the time-dependent brain graph evolution from a single observation in an end-to-end fashion.

\item \emph{On a methodological level.} EvoGraphNet is a unified prediction framework stacking a set of time-dependent graph GANs where the learning of the current gGAN in the cascade benefits from the learning of the previous gGAN. 

\item \emph{On clinical level.} EvoGraphNet can be used in the development of a more personalized medicine for the early diagnosis of brain dysconnectivity disorders.

\end{enumerate}

%%% ==========================================================================================
\section{Proposed Method}
%%% ==========================================================================================

In this section, we explain the key building blocks of our proposed EvoGraphNet architecture for time-dependent brain graph evolution synthesis from a single baseline timepoint. \textbf{Table}~\ref{tab:0} displays the mathematical notations used throughout our paper.

\begin{sidewaysfigure}
\includegraphics[width=19.5cm]{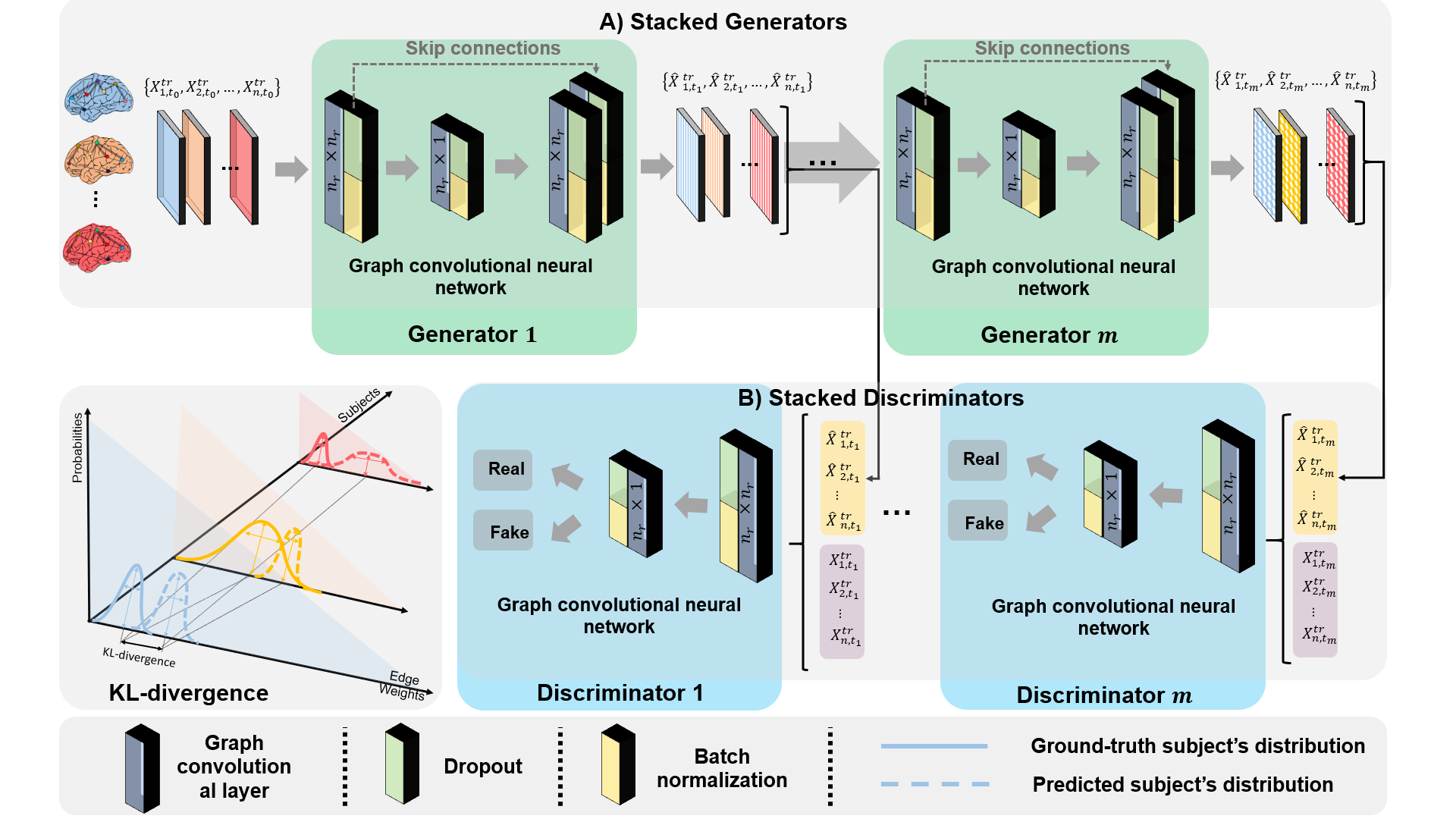}
\caption{\emph{Proposed EvoGraphNet architecture for predicting the longitudinal brain graph evolution from baseline timepoint $t_0$.} \textbf{(A)} \emph{Stacked generators network}. We develop a graph GAN (gGAN) that learns how to map an input brain graph measured at $t_{i-1}$ to a follow-up timepoint $t_i$. For $m$ timepoints, we design $m$ identical graph generators each composed of a three-layer graph convolutional neural network acting as an encoder-decoder network that mimics a U-net architecture with skip connections. For each $t_i, \ i >0$, each generator $G_i$ in the chain takes a set of predicted training brain graphs $\mathbf{\hat{X}}^{tr}_{t_{i-1}}$ by the previous generator and outputs a set of $\mathbf{\hat{X}}^{}_{t_{i}}$. \textbf{(B)} \emph{Stacked discriminator network}. For each generator $G_i$, we associate a discriminator $D_i$ aiming to distinguish the ground-truth brain graphs $\mathbf{X}^{tr}_{t_{i}}$ from the predicted brain graphs $\mathbf{\hat{X}}^{tr}_{t_{i}}$ by generator $G_i$. Hence, we design $m$ identical discriminators each composed of a two-layer graph convolutional neural network, each outputting a single score in the range of $[0, \dots, 1]$ expressing the \emph{realness} of $\mathbf{\hat{X}}^{tr}_{t_{i}}$. To enhance the generation quality at each follow-up timepoint $t_i$, we further integrate a Kullback-Leibler divergence loss  to better align the distribution of the predicted brain graphs with that of the ground-truth graphs.}
\label{mainfig}
\end{sidewaysfigure}

\begin{center}
\begin{table}
\caption{\label{tab:0} Major mathematical notations}
\resizebox{\textwidth}{!}{\begin{tabular}{c c}
 \hline
 \textbf{Mathematical notation}   & \hspace{1 cm} \textbf{Definition}  \\
 \hline
 $m$      &  number of timepoints to predict  \\ 
 $n_s$      &  number of training subjects  \\ 
 $n_r$    &  number of regions of interest in the brain \\
 $m_r$      &  number of edges in a brain graph   \\ 
 $\mu_k$ & mean of the connected edge weights for node $k$ in a ground-truth brain graph\\
 $\sigma_k$ & standard deviation of the connected edge weights for node $k$ in a ground-truth brain graph\\
 $\hat{\mu_k}$ & mean of the connected edge weights for node $k$ in a predicted brain graph\\
 $\hat{\sigma_k}$ & standard deviation of the connected edge weights for node $k$ in a predicted brain graph\\
 $p_k$ & normal distribution of the connected edge weights for a node $k$ in a ground-truth brain graph  \\
 $p_k$ & normal distribution of the connected edge weights for a node $k$ in the ground-truth brain graph  \\
 $\mathbf{X}_{t_i}^{tr}$ &  training brain graph connectivity matrices $\in\mathbb{R}^{n\times n_r \times n_r}$ at $t_i$ \\
 $\mathbf{\hat{X}}_{t_i}^{tr}$ &  predicted brain graph connectivity matrices $\in\mathbb{R}^{n\times n_r \times n_r}$ at $t_i$ \\
 $G_i$ & gGAN generator at timepoint $t_i$  \\
 $D_i$ & gGAN  discriminator  at timepoint $t_i$ \\
 $\mathcal{L}_{full}$ & full loss function \\
 $\mathcal{L}_{adv}$ & adversarial loss function \\
 $\mathcal{L}_{L_1}$ & $l_1$ loss function \\
 $\mathcal{L}_{KL}$ & KL divergence loss function \\
 $\lambda_1$ & coefficient of adversarial loss \\
 $\lambda_2$ & coefficient of $l_1$ loss \\
 $\lambda_3$ & coefficient of KL divergence loss \\
 $V$  & a set of $n_r$ nodes \\
 $E$  & a set of $m_r$ directed or undirected edges \\
 $l$  & index of layer \\
 $\mathbf{L}$  & transformation matrix $\in\mathbb{R}^{m_r\times s}$ \\
 $\mathcal{N}(k)$ & the neighborhood containing all the adjacent nodes of node $k$ \\
 $\mathbf{Y}^{l}(k)$ & filtered signal of node (ROI) $i$ $\in\mathbb{R}^{d_l}$ \\
 $\mathbf{F}^{l}_{k' k}$ & filter generating network \\
 $\omega^{l}$ & weight parameter \\
 $b^{l}$ & bias parameter \\
\hline
\end{tabular}}
\end{table}
\end{center}

\textbf{Overview of EvoGraphNet for time-dependent brain graph evolution prediction from a single source timepoint.} GANs \cite{Goodfellow} are deep learning models consisting of two  neural networks competing in solving a target learning task: a generator $G$ and a discriminator $D$. The generator is an encoder and decoder neural network aiming to learn how to generate fake samples that mimics a given original data distribution while the discriminator learns how to discriminate between the ground-truth data and the fake data produced by the generator. These two networks are trained in an adversarial way so that with enough training cycles, the generator learns how to better generate fake samples that look real and the discriminator learns how to better differentiate between ground-truth samples and generator's produced samples. We draw inspiration from the work of \cite{yang2020} which successfully translated a T1-weighted magnetic resonance imaging (MRI) to T2-weighted MRI using GAN and the work of \cite{stackgan2016} which proposed to use a stacked form of generators and discriminators for image synthesis. Therefore, as shown in \textbf{Fig.}~\ref{mainfig} our proposed EvoGraphNet is a chain of gGANs composed of $m$ generators and discriminators aiming to map a subject's brain graph measured at timepoint $t_0$ onto $m$ sequential follow-up timepoints. Excluding the baseline gGAN trained to map ground-truth baseline graphs at $t_0$ to the ground-truth brain graphs at timepoint $t_1$, a gGAN at time $t_i$ is trained to map the generated graphs by the previous gGAN at $t_{i-1}$ onto the ground-truth brain graphs at timepoint $t_i$.  Below is our adversarial loss function for one pair of generator $G_i$ and discriminator $D_i$ composing our gGAN at timepoint $t_i$: 
\begin{gather}
 arg min_{G_i} max_{D_i} \mathcal{L}_{adv}(G_i,D_i) =  \mathbb{E}_{G_i(\mathbf{X}_{t_i}) } [\log(D_i(G_i(\mathbf{X}_{t_i}))] +\mathbb{E}_{G_i(\mathbf{\hat{X}}_{t_{i-1}})} [\log(1 - D_i(G_i(\mathbf{\hat{X}}_{t_{i-1}}))] 
\end{gather}

$\hat{\mathbf{X}}_{t_{i-1}}$ denotes the predicted brain graph connectivity matrix by the previous generator $G_{i-1}$ in the chain, and which is inputted to the generator $G_{i}$ for training. $\mathbf{X}_{t_i}^{tr}$ denotes the target ground-truth brain connectivity matrix at timepoint $t_i$. $G_{i}$ is trained to learn a non-linear mapping from the previous prediction $\hat{\mathbf{X}}_{t_{i-1}}$ to the target ground truth $\mathbf{X}_{t_i}^{tr}$.

Since the brain connectivity changes are anticipated to be sparse over time, we further impose that the $l_1$ distance between two consecutive timepoints $t_{i-1}$ and $t_i$ to be quite small for $i \in \{1, \dots, m \}$. This acts as a time-dependent regularizer for each generator $G_i$ in our EvoGraphNet architecture. Thus, we express the proposed $l1$ loss for each subject $tr$ using the predicted brain graph connectivity matrix $\hat{\mathbf{X}}^{tr}_{t_{i-1}}$ by the previous generator $G_{i-1}$ and the ground-truth brain graph connectivity matrix to predict by generator $G_{i}$  as follows:

\begin{gather}
\mathcal{L}_{l1}(G_i,tr) = || \hat{\mathbf{X}}^{tr}_{t_{i-1}} - \mathbf{X}^{tr}_{t_{i}} ||_{1} 
\end{gather}

In addition to the $l1$ loss term introduced for taking into account time-dependency between two consecutive predictions, we further enforce the alignment between the ground-truth and predicted brain graph distribution at each timepoint $t_i$. Precisely,  we use the Kullback-Leibler (KL) divergence between both ground-truth and predicted distributions. KL divergence is a metric indicating the ability to discriminate between two given distributions. Thereby, we propose to add the KL divergence as a loss term to minimize the discrepancy between ground-truth and predicted connectivity weight distributions  at each timepoint $t_{i}$. To do so, first, we calculate the mean $\mu_k$ and the standard deviation $\sigma_k$ of connected edge weights for each node in each brain graph.

We define the normal distributions as follows:

\begin{gather}
p_k = N(\hat{\mu}_k, \hat{\sigma}_k)
\end{gather}
\begin{gather}
q_k = N(\mu_k, \sigma_k),
\end{gather}

where $p_k$ is the normal distribution for node $k$ in the predicted graph $\hat{\mathbf{X}}^{tr}_{t_{i}} = G_{i}(\hat{\mathbf{X}}^{tr}_{t_{i-1}})$ and $q_k$ is the normal distribution defined for the same node $k$ in $\mathbf{X}^{tr}_{t_{i}}$.\\

Thus, the KL divergence between the previously calculated normal distributions for each subject is expressed as follows:

\begin{gather}
\mathcal{L}_{KL}(t_i,tr) = \sum_{k=1}^{n_r} KL(p_k||q_k), 
\end{gather}

where each node's KL divergence is equal to:

\begin{gather}
KL(p_k||q_k) = \int_{-\infty}^{+\infty} p_k(x)log\frac{p_k(x)}{q_k(x)}dx
\end{gather}

\textbf{Full loss}. By summing up over all training $n_s$ subjects and all $m$ timepoints to predict, we obtain the full loss function to optimize follows:
\begin{gather}
\mathcal{L}_{Full} =  \sum_{i=1}^m \left( \lambda_1\mathcal{L}_{Adv}(G_i,D_i) + \frac{\lambda_2}{n_s} \sum_{tr=1}^{n_s}  \mathcal{L}_{l1}(G_i,tr) + \frac{\lambda_3}{n_s} \sum_{tr=1}^{n_s} \mathcal{L}_{KL}(t_i,tr)  \right)
\end{gather}

where $\lambda_1$, $\lambda_2$, and $\lambda_3$  are hyperparameters controlling the significance of each loss function in the overall objective to minimize.\\

\textbf{The generator network design}. As shown in \textbf{Fig.}~\ref{mainfig}--A, our proposed EvoGraphNet is composed $m$ generators aiming to predict the subject's brain graphs at $m$ follow-up timepoints. Since all  generators are designed identically and for the sake of simplicity,  we propose to detail the architecture of one generator $G_i$ aiming to predict subjects' brain graphs at timepoint $t_{i}$. \\

Our proposed generator $G_i$ consists of a three-layer encoder-decoder graph convolutional neural network (GCN) leveraging the dynamic edge convolution process implemented in \cite{SimonovskyK17} and imitating a U-net architecture \cite{Olaf:2015} with skip connections which enhances the decoding process with respect to the encoder's embedding. For each $t_i, \ i >0$, each generator $G_i$ in the chain takes a set of predicted training brain graphs $\mathbf{\hat{X}}^{tr}_{t_{i-1}}$ by the previous generator and outputs a set of $\mathbf{\hat{X}}^{}_{t_{i}}$.  Hence, our generator $G_i$ contains three graph convolutional neural network layers to which we apply batch normalization \cite{ioffe2015} and dropout \cite{xiao2016} to the output of each layer. These two operations undeniably contribute to simplifying and optimizing the network training. For instance, batch normalization was proven to accelerate network training while dropout was proven to eliminate the risk of overfitting.\\

\textbf{The discriminator network design}. We display the architecture of our discriminator in \textbf{Fig.}~\ref{mainfig}--B. Similar to the above sub-section, all discriminators share the same design, thus we propose to only detail one discriminator $D_i$ trained at timepoint $t_i$. We couple each generator $G_i$ with its corresponding discriminator $D_i$ which is also a graph neural network inspired by \cite{SimonovskyK17}. Our proposed discriminator is a two-layer graph neural network that takes as input a concatenation of the generator's output $\mathbf{\hat{X}}^{tr}_{ t_{i}}$ and the ground-truth subjects' brain graphs  $\mathbf{X}^{tr}_{ t_{i}}$. The discriminator outputs a value in the range of $[0, \dots, 1]$ measuring \emph{the realness} of the synthesized brain graph  $\mathbf{\hat{X}}^{tr}_{ t_{i}}$ at timepoint $t_i$. As reflected by our adversarial loss function, we design our gGAN's loss function so that it maximizes the discriminator's output score for $\mathbf{X}^{tr}_{ t_{i}}$ and minimizes it for $\mathbf{\hat{X}}^{tr}_{ t_{i}}$, which is the output of the previous generator in the chain $G_i(\mathbf{\hat{X}}^{tr}_{ t_{i-1}})$.\\

\textbf{Dynamic graph-based edge convolution}. Each of the graph convolutional layers of our gGAN architecture uses a dynamic graph-based edge convolution operation proposed by \cite{SimonovskyK17}. In particular, let $G=(V, E)$ be a directed or undirected graph where $V$ is a set of $n_{r}$ ROIs and $E  \subseteq V \times V $ is a set of $m_{r}$ edges. Let $l$ be the layer index in the neural network.  We define $\mathbf{Y}^{l}: V \rightarrow \mathbb{R}^{d_{l}}$ and $\mathbf{L}: E \rightarrow \mathbb{R}^{d_m}$ which can be respectively considered as two transformation matrices (i.e., functions) where $\mathbf{Y}^{l} \in \mathbb{R}^{n_{r} \times d_{l}}$ and $\mathbf{L} \in \mathbb{R}^{m_{r} \times d_m}$. $d_m$ and $d_{l}$ are dimensionality indexes. We define by $\mathcal{N}(k)= \left\{ k'; (k', k) \in E \right\} \cup \left\{k \right\}  $  the neighborhood of a node $k$ containing all its adjacent ROIs. 

The goal of each layer in each generator and discriminator in our EvoGraphNet is to output the graph convolution result which can be considered as a filtered signal $\mathbf{Y}^{l}(k) \in \mathbb{R}^{d_{l}}$ at node $k$'s neighborhood $k' \in \mathcal{N}(k) $. $\mathbf{Y}^{l}$ is expressed as follows:  

\begin{gather}
\mathbf{Y}^{l} (k) = \frac{1}{\mathcal{N}(k)} \sum_{k' \in \mathcal{N}(k)} \mathbf{\Theta}^{l}_{k' k} \mathbf{Y}^{l-1} (k') + b^{l}, 
\end{gather}

where $\mathbf{\Theta}^{l}_{k' k} = \mathbf{F}^{l} (\mathbf{L}(k', k); \omega^{l})$. We note that $\mathbf{F}^{l}: \mathbb{R}^{d_m} \rightarrow \mathbb{R}^{d_{l} \times d_{l}-1}$ is the filter generating network, $\omega^{l} $ and $b^{l}$ are model parameters that are updated only during training. 

% %% ***************************************************************************** %%
\section{Results and Discussion}
% %% ***************************************************************************** %% 

\textbf{Evaluation dataset.} We used $113$ subjects from the OASIS-2\footnote{\url{https://www.oasis-brains.org/}} longitudinal dataset \cite{Marcus:2010}. This set consists of a longitudinal collection of 150 subjects aged 60 to 96. Each subject has 3 visits (i.e., timepoints), separated by at least one year. For each subject, we construct a cortical morphological network derived from cortical thickness measure using structural T1-w MRI as proposed in \cite{Mahjoub:2018,Soussia:2018b,Lisowska:2017}. Each cortical hemisphere is parcellated into $35$ ROIs using Desikan-Killiany cortical atlas. We program our EvoGraphNet using PyTorch Geometric library \cite{Pytorchgeometric:2019}. 

 \textbf{Parameter setting.} In \textbf{Table}~\ref{tab:2}, we report the mean absolute error between ground-truth and synthesized brain graphs at follow-up timepoints $t_1$ and $t_2$. We set each pair of gGAN's hyperparameter as follows: $\lambda_1 = 2$,  $\lambda_2 = 2$, and $\lambda_3 = 0.001$. Also, we chose AdamW \cite{adamw} as our default optimizer and set the learning rate at $0.01$ for each generator and $0.0002$ for each discriminator. We set the exponential decay rate for the first moment estimates to $0.5$, and the exponential decay rate for the second-moment estimates to $0.999$ for the AdamW optimizer. Finally, we trained our gGAN for $500$ epochs using a single Tesla V100 GPU (NVIDIA GeForce GTX TITAN with 32GB memory).

\textbf{Comparison Method and evaluation.} To evaluate the reproducibility of our results and the robustness of our EvoGraphNet architecture to training and testing sample perturbation, we used a 3-fold cross-validation strategy for training and testing. Due to the lack of existing works on brain graph evolution prediction using geometric deep learning, we proposed to evaluate our method against two of its variants. The first comparison method (i.e., base EvoGraphNet) is an ablated of our proposed framework where we remove the KL divergence loss term while for the second comparison method (i.e., EvoGraphNet (w/o KL) $+$ Topology), we replace our KL-divergence loss with a graph topology loss. Basically, we represent each graph by its node strength vector storing the topological strength of each of its nodes. The node strength is computed by adding up the weights of all edges connected to the node of interest. Next, we computed the $L_2$ distance between the node strength vector of the ground-truth and predicted graphs. \textbf{Table}~\ref{tab:2} shows the MAE results at $t_1$ and $t_2$ timepoints.

\begin{table}[ht!]
\begin{center}

\caption{\label{tab:2} Prediction accuracy using mean absolute error (MAE) of our proposed method and comparison methods at $t_1$ and $t_2$ timepoints.}

\resizebox{\textwidth}{!}{\begin{tabular}{c | c | c | c | c}
\hline
 & \multicolumn{2}{c|}{$t_1$} & \multicolumn{2}{c}{$t_2$} \\
 \hline
 \textbf{Method}   
 &  \begin{tabular}{c c}
    \textbf{Mean MAE}   \\
     $\pm$ std
 \end{tabular} 
 & \begin{tabular}{c c}
     \textbf{Best}   \\
     \textbf{MAE}
 \end{tabular}
 &   \begin{tabular}{c c}
     \textbf{Mean MAE}   \\
     $\pm$ std
 \end{tabular} 
 &  \begin{tabular}{c c}
     \textbf{Best}   \\
     \textbf{MAE}
 \end{tabular} \\
 \hline
 Base EvoGraphNet (w/o KL) & $0.05626 \pm 0.00446$ & $\mathbf{0.05080}$ & $0.13379 \pm 0.01385$ & $0.11586$ \\
 EvoGraphNet (w/o KL) $+$ Topology & $0.05643 \pm 0.00307$ & $0.05286$ & $0.11194 \pm 0.00381$ & $0.10799$ \\
 EvoGraphNet & $\mathbf{0.05495 \pm 0.00282}$ & $0.05096$ & $\mathbf{0.08048 \pm 0.00554}$ & $\mathbf{0.07286}$ \\

 \hline
\end{tabular}}
\end{center}
\end{table}

Clearly, our proposed EvoGraphNet outperformed the comparison methods in terms of mean (averaged across the 3 folds) and the best MAE for prediction at $t_2$. It has also achieved the best mean MAE for brain graph prediction at $t_1$. However, the best MAE for prediction at $t_1$ was achieved by base EvoGraphNet which used only a pair of loss functions (e.i., adversarial loss and $l1$ loss). This might be due to the fact that our objective function is highly computationally expensive compared to the base EvoGraphNet making it less prone to learning flaws such as overfitting. Also, we notice that the prediction error at $t_1$ is lower than $t_2$, which indicates that further observations become more challenging to predict from the baseline timepoint $t_0$. Overall, our EvoGraphNet architecture with an additional KL divergence loss has achieved the best performance in foreseeing brain graph evolution trajectory and showed that stacking a pair of graph generator and discriminator for each predicted timepoint is indeed a promising strategy in tackling our time-dependent graph prediction problem.

\textbf{Limitations and future work.} While our graph prediction framework reached the lowest average MAE against benchmark methods in predicting the evolution of brain graphs over time from a single observation, it has a few limitations. So far, the proposed method only handles brain graphs where a single edge connects two ROIs. In our future work, we aim to generalize our stacked gGAN generators to handle brain hypergraphs, where a hyperedge captures high-order relationships between sets of nodes. This will enable us to better model and capture the complexity of the brain as a highly interactive network with different topological properties. Furthermore, we noticed that the joint integration of KL divergence loss and the topological loss produced a negligible improvement in the brain graph evolution prediction over time. We intend to investigate this point in depth in our future work. Besides, we aim to use a population network atlas as introduced in \cite{mhiri2020joint} to assist EvoGraphNet training in generating biologically sound brain networks.

% %% ***************************************************************************** %%
\section{Conclusion}
% %% ***************************************************************************** %%

In this paper, we proposed a first-ever geometric deep learning architecture, namely EvoGraphNet, for predicting the longitudinal evolution of a baseline brain graph over time. Our architecture chains a set of gGANs where the learning of each gGAN in the chain depends on the output of its antecedent gGAN. We proposed a time-dependency loss between consecutive timepoints and a distribution alignment between predicted and ground-truth graphs at the same timepoint.  Our results showed that our time-dependent brain graph generation framework from the baseline timepoint can notably boost the prediction accuracy compared to its ablated versions. Our EvoGraphNet is generic and can be trained using any given number of prediction timepoints, and thus can be used in predicting both typical and disordered changes in brain connectivity over time. Therefore, in our future research, we will be testing our designed architecture on large-scale connectomic datasets of multiple brain disorders such as schizophrenia by exploring the ability of the \emph{synthesized} brain graphs at later timepoints to improve neurological and neuropsychiatric disorder diagnosis from baseline.

% %% ***************************************************************************** %%
\section{Supplementary material}
% %% ***************************************************************************** %%

We provide three supplementary items for reproducible and open science: 

\begin{enumerate}
	\item A 6-mn YouTube video explaining how our prediction framework works on BASIRA YouTube channel at \url{https://youtu.be/aT---t2OBO0}.
	\item EvoGraphNet code in Python on GitHub at \url{https://github.com/basiralab/EvoGraphNet}. 
	\item A GitHub video code demo of EvoGraphNet on BASIRA YouTube channel at \url{https://youtu.be/eTUeQ15FeRc}. 
\end{enumerate}

% %% ***************************************************************************** %%
\section{Acknowledgement}
% %% ***************************************************************************** %%

This project has been funded by the 2232 International Fellowship for
Outstanding Researchers Program of TUBITAK (Project No:118C288, \url{http://basira-lab.com/reprime/}) supporting I. Rekik. However, all scientific contributions made in this project are owned and approved solely by the authors.

%%%%%%%%%%%%%%%%%%%%%%%%%%%%%%%%%%%%%%%%%%%%%%%%%%%%%%%%%%%%%%%%%%%%%%%%%%%%%%%%%%%%%%%%%%%%%%%%%%%%%%%%%%%%
\bibliography{Biblio}
\bibliographystyle{splncs}
\end{document}